\documentclass[aps,prd,showpacs,nofootinbib,superscriptaddress,preprint,tightenlines]{revtex4}

\usepackage{graphicx}
\usepackage[centertags]{amsmath}
\usepackage{amsfonts}
\usepackage{amssymb}
\usepackage{amsthm}
\usepackage{newlfont}
\usepackage{subfigure}
\usepackage{multirow}
\usepackage{accents}
\usepackage{color}
\usepackage{float}
\usepackage{leftidx}

\def\bea{\begin{eqnarray}}
\def\eea{\end{eqnarray}}

\usepackage{graphicx}

\def\be{\nopagebreak[3]\begin{equation}}
\def\ee{\end{equation}}
\def\ba{\nopagebreak[3]\begin{eqnarray}}
\def\ea{\end{eqnarray}}

\def\w{\omega}

\newcommand{\teta}{\rlap{\lower2ex\hbox{$\,\tilde{}$}}\eta{}}

\def \Gc{\Gamma_{\text{cov}}}

\newcommand{\W}{\mathbb{W}}
\newcommand{\V}{\mathbb{V}}
\newcommand{\U}{\mathbb{U}}
\newcommand{\Y}{\mathbb{Y}}
\newcommand{\om}{{\mathrm o}}   

\usepackage{enumerate}

\usepackage{colordvi}
\usepackage[centertags]{amsmath}
\usepackage{amsfonts}
\usepackage{amssymb}
\usepackage{amsthm}
\usepackage{dsfont}
\usepackage{newlfont}
\usepackage{hyperref}
\newcommand{\md}{{\mathrm d}}

\usepackage{ulem}

\newcommand{\ed}{\md \!\!\!\! \md\, }

\def\S{\Sigma}

\def\de{\delta}

\def\D{\Delta}

\def\eps{{}^2\!\epsilon}
\def\w{\wedge}
\def\co{8\pi G}

\def\La{\mathcal{L}}

\def\k{\kappa}

\newcommand{\pbi}[1]{{\underset{^\leftarrow}{{#1}}}} 
\newcommand{\WIHeq}{\overset{_\Delta}{=}}           

\newcommand{\A}{\mathcal{A}}
\newcommand{\sdA}{\leftidx{^+}{\mathcal{A}}}          
\newcommand{\sdF}{\leftidx{^+}{\mathcal{F}}}       

\begin{document}

\title{On covariant and canonical Hamiltonian formalisms: Weakly Isolated Horizons}
\author{Alejandro Corichi}\email{corichi@matmor.unam.mx}
\affiliation{Centro de Ciencias Matem\'aticas, Universidad Nacional Aut\'onoma de
M\'exico, UNAM-Campus Morelia, A. Postal 61-3, Morelia, Michoac\'an 58090,
Mexico}

\author{Juan D. Reyes}
\email{jdreyes@uach.mx}
\affiliation{Facultad de Ingenier\'\i a,
Universidad Aut\'onoma de Chihuahua, 
Nuevo Campus Universitario, Chihuahua 31125, Mexico}

\author{Tatjana Vuka\v{s}inac}
\email{tatjana.vukasinac@umich.mx}
\affiliation{Facultad de Ingenier\'\i a Civil, Universidad Michoacana de San Nicol\'as de Hidalgo, 
Morelia, Michoac\'an 58000, Mexico}

\begin{abstract}

The Hamiltonian description of classical gauge theories is a well studied subject. The two best known approaches, namely the {\it covariant} and {\it canonical} Hamiltonian formalisms, have received a lot of attention in the literature. The relation between them has also been extensively analyzed.
However, a complete understanding of this relation is still not available, specially for gauge theories that are defined over regions with boundaries.  
Here we consider this issue, for spacetimes with isolated horizons as inner boundaries (representing black holes in equilibrium), and assess whether their corresponding descriptions can be seen as equivalent. First, we review and reanalyze both formalisms from anew and show that, if we compare them at face value, there are  differences between  them, for instance in the derivation of horizon energy, a physical observable of the theory, as well as in the number of boundary degrees of freedom. Here we construct and compare possible extensions of the corresponding phase spaces, in a way that has not been explored before. We analyze different possible interpretations in both approaches and show how one can  achieve correspondence between results within covariant and canonical formalisms. 
Along the way, we shed light on the role of boundary terms in the symplectic structure in relation to boundary degrees of freedom.

\end{abstract}

\maketitle


\section{Introduction}
\label{sec:1}

Gravitational theories for spacetimes with a null internal boundary, modeling a black hole (BH)  horizon, have been intensely studied from different viewpoints, using diverse formalisms. In particular, Hamiltonian methods have been used to treat consistent action principles and provide a convenient description both at the classical level and with the purpose of quantization. The literature on the subject is vast and dates back several decades now. In order to treat BH-boundary conditions for an action principle, the formalism of  `isolated horizons' was introduced in \cite{ACKclassical,ABF}. The theory was further developed with both {\it{canonical}} and {\it{covariant}} Hamiltonian methods, which allowed to obtain an expression for the horizon mass and recover black hole mechanics. This was also the starting point for (loop) quantum gravity entropy calculations \cite{ABCK,ABK}.

At the classical level, both canonical and covariant Hamiltonian methods have been widely explored for gravitational theories at infinity \cite{Regge&Teitelboim,abr}, and also for BH-type inner horizons \cite{ACKclassical,ABF,AFK,ABL}. The relation between these two methods has received some recent attention \cite{Wald,Espanolitos,CRV-3}, specially when boundaries are present. The purpose of this manuscript is to explore this issue and directly compare both methods for a class of isolated horizons boundary conditions (a special class of so called Carrollian structures). The study of field theories with boundaries, be them timelike or null, has triggered a good deal of contributions in the literature, ranging from the study of edge states and observables \cite{Freidel}, general covariant Hamiltonian methods \cite{Harlow-Wu,Espanolitos2,Troessaert,review,Witten,crv1}, and identification of generic Hamiltonian dynamics at 
boundaries of the spacetime region \cite{B&F,B&F1,Sorkin,Wieland}. Our purpose is to continue with the study of  weakly isolated horizons (WIH) 
along the lines of \cite{ABF,AFK,ACKclassical,CRV-1,CRV-2}.

Here, we study the action for gravity in first order formalism. For simplicity, we consider spacetimes, with weakly isolated horizons as an internal boundary, using self-dual variables. 
This theory was studied for the first time in \cite{ACKclassical}, and \cite{ABF}, for strongly isolated horizons, using the canonical formalism. There it was shown that one could obtain a horizon contribution to the Hamiltonian that can be identified with a generalization of the Smarr mass formula \cite{Smarr}. In the following works, the covariant phase space formalism turned out to be more  convenient for the study of the laws of black hole mechanics \cite{AFK,ABL}. However,  these results refer to Palatini and Holst actions in real variables \cite{cg,Barbero&al}. It has also been shown that many properties, including the symplectic structure and conserved quantities for the theory in self-dual variables can be obtained as a special case of the Holst action, for $\gamma =i$ \cite{Sahlmann2023,CRV-2}.

The comparison between the two approaches, namely, the covariant and the canonical ones, raises the question of the compatibility of the WIH symplectic structures, 
reflected in the horizon degrees of freedom and residual gauge symmetry.
A question that, in our opinion, has not been clearly addressed in the literature. The purpose of this work is to try to fill in this gap.
 Here we review both formalisms and analyze to some extent their seemingly different descriptions. Our viewpoint in this manuscript is that one should compare not only the symplectic structures (as in  \cite{Wald,Espanolitos,CRV-3}), but also and most importantly possible physical observables. In the case of non-rotating WIH, the horizon energy (and mass) turns out to be a physical observable, that is, a gauge invariant function on the phase space of the theory (projectable to the reduced phase space in both approaches). Thus, it provides an appropriate object for the comparison of both formalisms, without the need of going to the reduced phase space explicitly.
 
As we shall discuss in detail, in the covariant phase space formalism, in the process of
constructing the (conserved) symplectic structure, one needs to introduce two potentials ($\psi_R$ and $\psi_I$).
These potentials contribute to a WIH boundary term in the symplectic structure. At  first sight, it would seem that one is introducing new degrees of freedom on the boundary and thus extending the original covariant phase space. However, these potentials are determined by the pullbacks to the horizon of the bulk fields (namely, the connection), and can be completely fixed by an appropriate choice of initial conditions, as has been done, for example, in \cite{AFK,ABL,cg}.
As a result, though they appear in the symplectic structure, the potential fields themselves are non-dynamical.
It turns out that Hamiltonian evolution can be properly defined, with a corresponding Hamiltonian function and its variation giving the  first law of black hole dynamics.
Nevertheless, the presence of horizon boundary terms in the symplectic structure, strongly suggests one may also obtain consistent descriptions by reinterpreting either or both of the potential fields as true or independent boundary degrees of freedom in an extended phase space. These options have not been previously studied and here we explore these possibilities.

On the other hand, in the canonical formalism, it had been shown  that there are several  formulations, corresponding to the introduction of $\psi_R$ or $\psi_I$ as boundary degrees of freedom, that do not always provide the expected horizon mass as a horizon boundary term in the Hamiltonian \cite{CRV-2}. Here we shall reanalyze these cases and show how one may obtain a consistent description. On the contrary to the covariant phase space approach, the necessity to introduce the potentials in the canonical formalism, 
is more subtle. The natural choice from the canonical perspective, is to treat  boundary terms in the action that do not contain time derivatives, as contributing to the Hamiltonian boundary term.  As it turns out, this choice does not lead to a consistent formulation preserving all degrees of freedom and giving the correct value of mass. One boundary term must be interpreted as a kinetic term by introducing the non-dynamical potential $\psi_I$. Thus, the canonical symplectic structure must have a boundary term too.  

If additionally one wants to compare both formalisms, one needs to introduce $\psi_R$ as a non-dynamical field in the canonical formalism as well. Furthermore, here one can also explore the possibility  to extend the original phase space, considering one of the potentials or both of them as dynamical. In all these cases, the potentials contribute to the canonical symplectic structure. The horizon mass can be obtained for the appropriate (physical) boundary Hamiltonian flow. As we will show, these  canonical cases correspond in a precise way to the original covariant formulation and covariant extensions alluded before.

After the analysis of these various possibilities in both formalisms, the main lesson is that in this case, one does have complete correspondence between covariant and canonical phase space descriptions, without extending the original phase spaces. However, one needs to introduce non-dynamical fields on the WIH boundary. 
Furthermore, the equivalence of both descriptions can also be achieved after  appropriate extensions of the original phase spaces, by reinterpreting $\psi_R$ and $\psi_I$ as a dynamical fields on the WIH.
These fields, or more fundamentally the presence of a boundary term in the symplectic structure, are what make the comparison less than straight forward, due to the subtleties and ambiguities they introduce. Here we explore the interplay between both formalisms and show how one can correctly extend the original phase spaces.

In the following, we will expose the details of our claims. The paper is organized as follows. We present some preliminaries regarding both formalisms in Sec. \ref{sec:2}.
In Sec. \ref{sec:3} we recall the definition of a WIH and its main properties. Next, in Sec. \ref{sec:4} we explore in detail the relation between covariant and canonical WIH symplectic structures and horizon Hamiltonians, for gravity in self-dual variables. Section \ref{sec:5} is dedicated to conclusions and further discussion of our results.

\section{Basic structures of covariant and canonical Hamiltonian analysis}
\label{sec:2}

In this section we will give a very short reminder of the basic ideas of both approaches, based mostly on \cite{Witten,abr,Wald,CRV2} for the covariant case, and \cite{HT,Troessaert,CV-M+P,GNH,Barbero1,Barbero2} for the canonical one, among many others.
For simplicity and concreteness here we shall consider first order gauge theories, whose configuration space is formed from $1-$forms $\mathbf{A}^I(x)$, where $I$ denotes all internal indices. The covariant action is given on a spacetime region $\mathcal{M}$ with boundary, $\partial\mathcal{M}=\Sigma_1\cup\Sigma_2\cup\Delta\cup\tau_\infty$, where $\Sigma_{1,2}$ are two (arbitrary) Cauchy surfaces, $\Delta$ is a WIH horizon and $\tau_\infty$ is an asymptotic time-like hypersurface. The fields $\mathbf{A}^I(x)$ should satisfy appropriate boundary conditions,  that we shall specify in the next section.

The covariant phase space $(\Gamma_{\text{cov}},\pmb{\omega})$,  consists of the space of solutions to the equations of motion, that satisfy given boundary conditions, $\Gamma_{\text{cov}}$, equipped with a pre-symplectic structure $\pmb{\omega}$, that is a degenerate, closed $2-$form on $\Gamma_{\text{cov}}$. 
The degeneracy of $\pmb{\omega}$ is deeply related to the invariance of the theory under gauge transformations.  

On the other hand, the canonical phase space $(\Gamma_{\text{can}},\Omega)$ is the space of all allowed initial data on a Cauchy hypersurface, $\Gamma_{\text{can}}$, with a
non-degenerate, closed $2-$form $\Omega$, the kinematical symplectic structure.
In gauge theories there are first class constraints and when we restrict to the constraint surface, the pullback of the symplectic structure $\bar{\Omega}$ becomes degenerate. Then, one can compare $\pmb{\omega}$ with $\bar{\Omega}$, in the precise sense given in \cite{CRV-3}. It turns out that they do not always coincide.

Our main interest is the WIH symplectic structure and the corresponding Hamiltonians in both approaches, and in the following we shall first recall both constructions.

\subsection{Covariant Hamiltonian analysis}
\label{subsec:2.1}

Let us start from a generic,  first order covariant action, without boundary terms,
\begin{equation}
S[\mathbf{A}] = \int_{\mathcal{M}}  \mathbf{L}\, .\label{CovAction}
\end{equation}
Its variation can be written as 
\begin{equation}\label{VarActFormsWithoutBoundary}
\ed S[\delta] := \delta S [\mathbf{A}]   = \int_{\mathcal{M}} {\mathbf{E}}_I \wedge\delta {\mathbf{A}}^I  + \int_{\mathcal{M}} \md \theta ( {\mathbf{A}}^I,\delta {\mathbf{A}}^I).
\end{equation}
where $\, \ed$ denotes the exterior derivative on the covariant phase space and the  variations of the fields are identified with tangent vectors to $\Gamma_{\text{cov}}$.
The second term of the RHS is called the symplectic potential\footnote{Sometimes, one finds in the literature that $\Theta$ is defined as an integral over $\Sigma$ and the symplectic structure $\pmb{\omega} =\ed\Theta$ is directly obtained as an integral over $\Sigma$. Here, we want to emphasize that one should be careful in order to obtain a preserved symplectic structure \cite{abr,CRV2}.}, and it can be written as,
\begin{equation}\label{SympPotentialWhitoutBoundary}
\Theta (\delta {\mathbf{A}}^I) := 
\int_{\partial \mathcal{M}} \theta ({\mathbf{A}}^I,\delta {\mathbf{A}}^I )\, .
\end{equation}
It is obtained after integration by parts and has a twofold role. First, it has to vanish, for given boundary conditions in order to have well defined variational principle that permits one to obtain the Euler-Lagrange equations of motion (EOM),  ${\mathbf{E}}_I=0$, in the bulk. In general, it may not vanish and one may have to add  an appropriate boundary term to the action, in order to make it differentiable. In particular, the counter term needed at infinity to make the self-dual action (\ref{SDactionVer2}) differentiable will play no role in our analysis.

On the other hand, $\Theta$ is also a starting point for the construction of a symplectic structure of the theory.  The exterior derivative of the symplectic potential, 
acting on tangent vectors $\delta_{1}$ and $\delta_{2}$ at an arbitrary point in  $\Gamma_{\text{cov}}$ is given by
\begin{equation}\label{edTheta}
\ed \Theta (\delta_{1}, \delta_{2}) 
= 2 \int_{\partial \mathcal{M}} \delta_{[1} 
\theta (\delta_{2]}):=\int_{\partial \mathcal{M}} J(\delta_{1}, \delta_{2})\, ,
\end{equation}
where $J(\delta_{1}, \delta_{2})$ is a spacetime $3-$form, the symplectic current.

Since on the space of solutions, $\ed S (\delta) = \Theta (\delta)$, we obtain
\begin{equation}
0 = \ed ^{2} S (\delta_{1}, \delta_{2}) =  \ed \Theta (\delta_{1}, \delta_{2}) = \left(-  \int_{\Sigma_{1}} + \int_{\Sigma_{2}}    + 
\int_{\Delta} + \int_{\tau_\infty} \right)  J \, .
\end{equation}
Where we now consider the boundary $\partial \mathcal{M}=\Sigma_1\cup\Sigma_2\cup\Delta\cup\tau_\infty$, consisting of `initial' and `final' Cauchy hypersurfaces $\Sigma_1$, $\Sigma_2$, and an `inner'  and `outer' boundaries, $\Delta$ and $\tau_\infty$, corresponding to a WIH and infinity respectively.
The signs are determined taking into account the orientation of the boundary with respect to the outer normals to the components of the boundary that appear in the Stokes' theorem. 

We are going to anticipate the results from the following section, where the integral over $\tau_\infty$ can be neglected \footnote{This term has to be combined with a contribution from the boundary term that needs to be added to the action at $\tau_\infty$ to make it differentiable \cite{ABF,CRV-1}. The sum of these terms vanishes and we are left only with a contribution from $\Delta$.},  and  $J=\md j$ on $\Delta$, leading to $\int_{\Delta}  J= (-\int_{S_{\Delta 1}}+\int_{S_{\Delta 2}}) j$, where $S_\Delta =\Sigma\cap\Delta$. Then, we can define the conserved pre-symplectic structure (independent of the Cauchy surface) as
\begin{equation}\label{CovSymplStructBT}
\pmb{\omega} (\delta_{1}, \delta_{2}) = \int_{\Sigma}  J  (\delta_{1}, \delta_{2})+\int_{S_\Delta}  j  (\delta_{1}, \delta_{2}) \, .
\end{equation}
In the following sections we shall analyze in detail the boundary term in (\ref{CovSymplStructBT}), and show that it does not necessarily imply that there are dynamical degrees of freedom on $\Delta$.

Since we are dealing with a gauge theory, this construction leads to a pre-symplectic structure, that has degenerate directions, $Y_i\in T\Gamma_{\text{cov}}$, such that $\pmb{\omega} (Y_i, X)=0$, for every $X\in T\Gamma_{\text{cov}}$.

In the covariant phase space framework, dynamical or `time' evolution may be conceived as a mapping between histories or points on phase space. It is then natural to require that the infinitesimal generator of such mappings or time evolution be a hamiltonian vector field $\delta_t$ on covariant phase. Therefore, the hamiltonian $H_t$ generating this evolution is determined (up to an additive constant) from
\be
\delta H_t :=\ed H_t (\delta ) = \pmb{\omega}(\delta , \delta_t)\, ,
\ee 
with $\delta_t{\mathbf{A}}^I={\mathcal{L}}_t{\mathbf{A}}^I$ the infinitesimal generator in the bulk, for a fixed vector field $t^\mu$ on each spacetime. We will see that on $\Delta$, $\delta_t$ has to be defined  with care, taking into account the  boundary conditions.


\subsection{Canonical Hamiltonian analysis}
\label{subsec:2.2}
In order to arrive at the canonical phase space $\Gamma_{\mathrm{can}}$ one again
starts from a covariant action (\ref{CovAction}), defined in a spacetime region 
$\mathcal{M}=\mathcal{I} \times  \Sigma$, where $ \mathcal{I}$ is a closed interval, and $\Sigma$ a Cauchy surface, defined by $t=\text{const}$. 
For 3+1 decomposition of the action one chooses a time evolution vector field $t^\mu$, such that $t^\mu\nabla_\mu t=1$. Furthermore, spacetime fields ${\mathbf{A}}^I_\mu$ are split into tangential (spatial), ${\mathbf{A}}^I_a$, and normal, $\phi^I := t^\mu {\mathbf{A}}^I_\mu$, components with respect to this foliation. In these variables the covariant action can be rewritten in its  canonical form:
\be
S_{\text{can}}
=\int_\mathcal{I} \bigl( P[{\mathcal{L}}_t {\mathbf{A}}^I] + P[{\mathcal{L}}_t \phi^I]
- H_{\text{C}} \bigr)\, \md t\, ,
\ee
where $P[{\cal{L}}_t\mathbf{A}] + P[{\mathcal{L}}_t \phi^I]$ is the kinetic term and $H_{\text{C}}$ is the canonical Hamiltonian of the theory. As we shall see in the continuation, in our case, both of them can acquire boundary terms. In general, the kinetic term is of the form \cite{CV-M+P,CRV-3},
\be
P[{\mathcal{L}}_t {\mathbf{A}}^I] + P[{\mathcal{L}}_t \phi^I] = \int_\Sigma\,  {\mathbf{P}}_I \wedge {\mathcal{L}}_t {\mathbf{A}}^I + \int_\Sigma\,  {\mathbf{P}}_{\phi I} \wedge {\mathcal{L}}_t \phi^I +
\int_{\partial\Sigma}\, {\pmb{\pi}}_i\wedge {\mathcal{L}}_t {\pmb{\alpha}}^i\, , 
\ee
Here $({\mathbf{A}(x)}^I,{\mathbf{P}}_I(x)); \phi^I,  {\mathbf{P}}_{\phi I}) $ are the  bulk canonical variables and
$({\pmb{\alpha}}^i(y),{\pmb{\pi}}_i(y))$ may be interpreted as boundary degrees of freedom \footnote{As wee shall show, this interpretation will depend on the exact definition of these fields.
}, where $i$ refers to all internal indices for the fields on $\partial\Sigma$, that in general are not the pullbacks of the fields on $\Sigma$. 
 Thus, one has obtained coordinates for the cotangent bundle $T^*\cal{C}=:\Gamma_{\mathrm{can}}$.

The kinetic term determines the kinematical non-degenerate symplectic structure of the theory that can also have a boundary term
\be
\Omega (\delta_1,\delta_2)= 2\int_\Sigma 
\bigl( \delta_{[1}{\mathbf{P}}_I\wedge\delta_{2]}{\mathbf{A}}^I   
+ \delta_{[1}{\mathbf{P}}_{\phi I}\wedge\delta_{2]}\phi^I \bigr)
+ 2\int_{\partial\Sigma} \delta_{[1}{\pmb{\pi}}_i\wedge\delta_{2]}{\pmb{\alpha}}^i\, .
\ee
In gauge theories there are first class constraints (FCC), $\mathbf{C}_k\approx 0$, $i=1,\dots,n$, and in general, there can be bulk and boundary ones. The theory can also have second class constraints (SCC), $\mathbf{D}_k\approx 0$, $k=1,\dots,m$. The pullback of $\Omega$ to the corresponding submanifold defined by the vanishing of the
SCC constraints, is {\it non-degenerate}. 
For our purposes, it is sufficient to restrict our considerations to the subspace of the phase space where the SCC constraints are imposed as strong equalities, $\Gamma_D\subset\Gamma_{\text{can}}$, and consider the pair $(\Gamma_D,\Omega_D)$ as a new phase space where dynamics is unfolding \cite{CV-M+P}. In our case, we will see that there are no SCC, so we will focus on first class constraint subspace.

The evolution should be tangent to the first class constraint surface,
$\bar{\Gamma}_{\text{can}}\subset\Gamma_{\text{can}}$ and the  pullback of $\Omega$ to $\bar{\Gamma}_{\text{can}}$ is degenerate. The Hamiltonian that governs the dynamics of the theory is not unique, it contains a linear combination of  FCC, and should satisfy 
\be
\ed H_{\text{C}} (Y)= \Omega (Y ,X_H)\, ,\label{HVF1}
\ee
where $X_H$ is the corresponding Hamiltonian vector field (HVF), that is unique for every particular choice of the multipliers, and $Y\in T\Gamma_{\text{can}}$. 
Similary, one can define the HVF, $X_{C_k}$ corresponding to smeared FCC.   

In the case when $\Omega$ has vanishing contribution from the boundary there cannot be any boundary terms in $\,\ed H_{\text{C}} (Y)$ and $\,\ed C_k (Y)$, and that imposes some boundary conditions on bulk canonical variables.  This corresponds to the standard Regge-Teitelboim scenario \cite{Regge&Teitelboim}. When the boundary conditions are given a-priori,  they might not be consistent with the vanishing of the above mentioned boundary terms and one needs to add counterterms to $H_{\text{C}}$ and $C_k$. 
In the case when $\Omega$ has non-vanishing contribution from the boundary, there is a boundary contribution to Hamiltonian vector fields, and generally there are also boundary conditions on the bulk configurations \cite{CV-M+P}. Nevertheless, in some theories $\,\ed H_{\text{C}} (Y)$ and $\,\ed C_k (Y)$ contain terms non-compatible with boundary terms in $\Omega$, and it is necessary to introduce counterterms. In general, the boundary term in $\Omega$ defines the boundary HVF, that gives  EOM on the boundary for ${\pmb{\alpha}}^i$ and ${\pmb{\pi}}_i$.

\subsection{Comparison of the two approaches}

As we have already mentioned, one can compare two Hamiltonian descriptions at the level of the symplectic structures and/or observables of the theory. Let us formulate these comparisons with more detail.

We have described how one can construct a pre-symplectic structure $\pmb{\omega}$ in $\Gamma_{\text{cov}}$ and the pullback $\bar{\Omega}$ of the canonical symplectic structure to $\bar{\Gamma}_{\text{can}}$. As explained in more detail in \cite{CRV-3,Espanolitos2},  one can define a \textit{canonical projection} $\bar{\Pi}: \Gamma_{\text{cov}}\to \bar{\Gamma}_{\text{can}}$, that assigns to each solution $s$ in $\Gc$ its `instantaneous' value $d$ at some `initial' hypersurface $\Sigma_0$.   The pullback $\tilde{\Omega}(s) :=\bar{\Pi}_*\bar{\Omega}(d)$ 
defines a pre-symplectic structure on $\Gamma_\text{cov}$. As shown in \cite{CRV-3,Espanolitos2}, the induced $\tilde{\Omega}(s)$ does not necessarily coincide with the naturally defined $\boldsymbol{\omega}(s)$ coming from the covariant phase space formalism. The pre-symplectic structures are $\bar\Pi$-equivalent if 
\begin{equation}
\tilde{\Omega}(s)=\boldsymbol{\omega}(s) \label{LWequivalence}
\end{equation}

As we will show in continuation, the direct comparison of the WIH symplectic structures is not straightforward. Instead, we shall focus on the comparison between Hamiltonians and the corresponding Hamiltonian flows on WIH, as well as the boundary degrees of freedom in both formalisms. The symplectic structures are fundamental for this analysis, but in a way they can be seen as auxiliary constructions. We will show in what cases one can obtain the correct expressions for the horizon energy. Furthermore, we will show that in the canonical formalism, in some cases it is necessary to modify the original 'natural' Hamiltonian flow on the WIH, in order to coincide with the flow obtained in the covariant phase space approach, that will permit us to obtain the correct expresion for the energy, extending the results of \cite{CRV-2}.

In the following sections we shall revise and compare the results of both approaches in the case of the theory of gravity in first order formalism in self-dual variables. We are interested in asymptotically flat spacetimes with a weakly isolated horizon as an internal boundary. 
As a first step,  we shall recall definitions and some of the basic properties of isolated horizons, see for instance, \cite{AFK} and \cite{CRV-2}.


\section{Isolated Horizons}
\label{sec:3}

Black holes in equilibrium (for instance some time after gravitational collapse or a merger), may be modeled by \textit{isolated horizons}. These are
non-expanding null hypersurfaces on $\mathcal{M}$. A 3-dimensional null hypersurface $\Delta$ has an induced metric $h_\pbi{\mu\nu}$ that is degenerate of signature $(0,+,+)$ \footnote{Following previous conventions, we will denote forms
pulled back to $\Delta$ using indexes with arrows under them.}.
This implies that any vector $\ell^\mu$ normal to $\Delta$ is null and satisfies the geodesic equation
\[
\ell^\mu\nabla_\mu\ell^\nu=\kappa_{(\ell)}\ell^\nu\,,
\]
and the non-affinity parameter $\kappa_{(\ell)}$ 
corresponds to surface gravity when one specializes to isolated horizons. 

For null hypersurfaces, it is convenient to use the null normal $\ell^\mu$  at each point $p\in\Delta$ to construct a Newman-Penrose null basis  $(k,\ell,m,\bar{m})$ on $T_p\mathcal{M}$. 
 A  null direction $k^\mu$ is transverse to $T_p\Delta$ on the light cone, normalized such that $k\cdot\ell=-1$. The orthogonal complement 
 of the plane spanned by $k^\mu$ and $\ell^\mu$ is a two dimensional spatial subspace of $T_p\Delta$, 
 where one can choose a  null basis given by a complex vector $m^\mu$ and its conjugate $\bar{m}$, such that $m\cdot\bar{m}=1$.  
We will call a basis constructed in this way a null basis adapted to $\Delta$.
It is not unique,
any two adapted null bases are related by 
a Lorentz rotation preserving the direction of $\ell^\mu$. 
The cross sectional area two-form 
$\leftidx{^2}{\epsilon}{}:=im\wedge\bar{m}\,$,
is invariantly defined on $\Delta$, in the sense that it is the same regardless of the choice of adapted null basis.
Similarly, the (two dimensional spatial) \textit{cross sectional metric}  
$2\,m_{(\pbi{\mu}}\bar{m}_{\pbi{\nu})}=h_\pbi{\mu\nu}\,$
is invariantly defined on $T_p\Delta$ and equal to the pullback metric.

The null vector $k^\mu$ allows one to define a projector to $T_p\Delta$, 
$
\Pi^\mu_{\;\;\nu}:=\delta^\mu_{\;\;\nu}+k^\mu\ell_\nu
$.
The second fundamental form of $\Delta$ with respect to $\ell^\mu$ 
is defined as
$
\Theta_{\mu\nu}:=\Pi^\sigma_{\;\;\mu}\Pi^\rho_{\;\;\nu}\nabla_\sigma\ell_\rho\,, 
$
and encodes the `kinematics' of the geodesic congruence of null generators of $\Delta$ with velocities $\ell^\mu$. Its trace $\theta_{(\ell)}:=g^{\mu\nu}\Theta_{\mu\nu}$ defines the \textit{expansion} of the congruence.

To model equilibrium horizons resulting from gravitational collapse, one first restricts the topology of $\Delta$ and incorporates the notion that the null geodesic generators should be non-expanding.
A null hypersurface $\Delta\subset\mathcal{M}$ is called a \textit{non-expanding horizon} (NEH) if it is topologically $S^2\times\mathbb{R}$,  
the expansion vanishes $\theta_{(\ell)}=0$, and  on $\Delta$, Einstein's equations hold and the stress-energy tensor $T_{\mu\nu}$ of matter satisfies the null dominant energy condition.

These conditions imply  that  on $\Delta$ 
the complete second fundamental form $\Theta_{\mu\nu}$ vanishes \cite{AFK,GJreview}. It follows that on a NEH, there exists a 1-form $\omega_\mu$, called \textit{rotation 1-form} in the context of black hole horizons, such that
\begin{equation} \label{rotationForm}
\nabla_{\pbi{\mu}}\ell^\nu\WIHeq\omega_\pbi{\mu}\ell^\nu\,,
\end{equation}
where we used the standard notation $\WIHeq$ for equalities valid only on $\Delta$. Then,
$\kappa_{(\ell)}\WIHeq\omega_\mu\ell^\mu$.
Equation (\ref{rotationForm}) also implies that the induced metric on $\Delta$ and the transverse area 2-form are Lie dragged along $\ell^\mu$,  
$\mathcal{L}_{\ell}\,h_{\pbi{\mu\nu}}\WIHeq 0$ and $\mathcal{L}_{\ell}\,\leftidx{^2}{\epsilon}{}\WIHeq 0$.

In a general null hypersurface $\Delta$ each choice of $k^\mu$ defines 
a (torsion-free) induced connection $\widehat{\nabla}_\pbi{\mu}$ compatible with the induced metric: $\widehat{\nabla}_\pbi{\mu}h_{\pbi{\nu\rho}}=0$.
While the NEH conditions guarantee a compatible connection is unique (independent of this choice),
this connection is not fully determined by the induced metric on $\Delta$ and hence contributes dynamical degrees of freedom independent of those of the metric for the geometry of a NEH.

Along with  $\omega_\pbi{\mu}$, on a NEH there is another relevant intrinsic one-form arising from the connection on $\Delta$, the transverse connection potential $V_\pbi{\mu}$. 
Using an adapted null basis, it is  defined as:
\begin{equation} \label{tV}
V_\pbi{\mu}:=\bar{m}_\nu\widehat{\nabla}_\pbi{\mu}m^\nu\,.
\end{equation}
and transforms as a $U(1)$-connection under $U(1)$-rotations that preserve $\ell$.


A \textit{weakly isolated horizon} $(\Delta,[\ell])$ is a non-expanding horizon where an equivalence class of null normals $[\ell]$ satisfying  
\begin{equation} \label{WIHcondition}
\mathcal{L}_{{\ell}}\,\omega_\pbi{\mu}\WIHeq 0  \qquad \text{ for all } \; \ell^\mu\in[\ell]
\end{equation}
has been singled out. Two normals $\tilde{\ell}\sim\ell$ belong to the same equivalence class iff $\tilde{\ell}^\mu=c\ell^\mu$ for some constant $c>0$ on $\Delta$.  
This symmetry condition or time-independence of a part of the induced connection
 is sufficient to ensure that the surface gravity $\kappa_{(\ell)}$ is constant on a WIH, that constitutes the zeroth law of black hole mechanics.
Every NEH can be made into a WIH by  appropriate choice of equivalence class $[\ell]$.  

Weakly isolated horizons may be classified according to their symmetries. In this work we restrict our analysis to non-rotating type I and III WIH's. Axi-symmetric horizons require somewhat different boundary conditions for the evolution fields $t^\mu$ and will no be treated here.


\section{Gravity in self-dual variables: Covariant and canonical approaches}
\label{sec:4}


This is the main section of the manuscript, where we present covariant and canonical Hamiltonian descriptions of vacuum general relativity in spinorial formalism for asymptotically flat spacetimes with a weakly isolated horizon as an internal boundary. We start with a brief  review of the spinorial formalism, more details can be consulted in \cite{ABF0, ABF, ACKclassical, CRV-2}. 
Then we construct the covariant phase space and the corresponding Hamiltonian as an observable that generates time evolution. We will focus our attention on WIH boundary contributions in both approaches.  In the following we recall and extend the results of the canonical analysis \cite{CRV-2} and compare them with the ones obtained in the covariant approach.

\subsection{Gravity action in self-dual variables}

The theory of general relativity can be rewritten in many different ways depending on the election of configuration variables. Here we are interested in its spinorial description, where the basic variables are a pair of one-forms $(\sigma^{AA'},\leftidx{^+}{\A}{_{AB}})$, see, for example \cite{AshtekarLectures,ACKclassical, ThiemannBook}. The internal primed and unprimed indices are $SL(2,\mathbb{C})$ spinor indices,  where $\sigma^{AA'}$ is a soldering form for primed and unprimed spinors and $\leftidx{^+}{\A}{_{AB}}$ is a $SL(2,\mathbb{C})$ self-dual connection.

The soldering form is anti-hermitian and defines a real Lorentzian metric, $g_{\mu\nu}=\sigma_\mu^{AA'}\sigma_{\nu AA'}$. The self-dual connection defines a derivative operator on unprimed spinors as, $DX_A=\partial X_A+ \leftidx{^+}{\A}{_{A}^{\;\;B}}X_B$ \cite{ACKclassical}.

For convenience, one can introduce a fixed dyad basis $(\om^A,\iota^A)$ in spinor space $\mathbb{C}^2$ at each point, such that  
$\iota^A\iota_A=0\, , \  \om^A\om_A =0\, , \ \iota^A\om_A=-\iota_A\om^A=1$,
and similarly for the conjugate basis $(\bar{\om}^{A'},\bar{\iota}^{A'})$. 
The soldering form, in the adapted null basis, is expanded as \cite{CRV-1}:
\begin{equation}  \label{SL2CsolderingExpansion}
\sigma^\mu_{AA'}=-i\, k^\mu \om_A\bar{\om}_{A'}-\, \ell^\mu \iota_A\bar{\iota}_{A'} + i\,\bar{m}^\mu \om_A\bar{\iota}_{A'} + i\, m^\mu\iota_A\bar{\om}_{A'} \, .
\end{equation}
On $\Delta$ one can define 
the spin connection compatible with the soldering form or vector, $\nabla_\mu\sigma^\nu_{AA'}=0$,
and write down an expansion for the self-dual connection potential in terms of the one-forms $(\W ,\V ,\U ,\Y)$ and the dyad $(\iota_A,\om_A)$ \cite{CRV-1}:
\begin{equation}\label{SDAexpansion}
\leftidx{^+}{\A}{_{AB}}=-(\W+\V)\,\iota_{(A}\om_{B)}+\bar{\U}\,\om_A\om_B-\Y\,\iota_A\iota_B\, ,
\end{equation}
where
$\W_\mu:=-k_\nu\nabla_\mu \ell^\nu\,, \ \ \V_\mu:=\bar{m}_\nu\nabla_\mu m^\nu\,, \ \ \U_\mu:=m_\nu\nabla_\mu k^\nu\,, \ \ \Y_\mu:=m_\nu\nabla_\mu \ell^\nu
$.
Expansion (\ref{SL2CsolderingExpansion}) is valid on the bulk as well as on $\Delta$, while the expansion (\ref{SDAexpansion}) is valid only on $\Delta$ (where we can use the compatibility condition).

The first order covariant action for gravity in self-dual variables, is given by 
(see, for example, \cite{Samuel,JacobsonSmolin,AshtekarLectures,ACKclassical,ABF})\footnote{Up to the multiplicative constant $\frac{1}{\co}$, that we ignore, for simplicity.}
\begin{align}
 S_{\text{SD}}(\Sigma,\sdA) =-i\int_{\mathcal{M}} \, \Sigma^{\;AB}\wedge\sdF_{\,AB}  \, ,     \label{SDactionVer2}
  \end{align}
where the two-forms $\Sigma^{\;AB}:= \sigma^{\;AA'}\wedge\sigma_{\;A'}\,^{B}$ are (spacetime) self-dual \cite{ACKclassical}, and
\be
\sdF_{\mu\nu}^{\,AB}=2\,\partial_{[\mu}\sdA_{\nu ]}^{\;AB}+\sdA_{[\mu\ C}^{\ A}\sdA_{\nu ]}^{\,CB}\, ,
\ee
is the curvature of the self-dual connection. The  action is defined on a spacetime region $\mathcal{M}$ with boundary, $\partial\mathcal{M}=\Sigma_1\cup\Sigma_2\cup\Delta\cup\tau_\infty$, where, as before, $\Sigma_{1,2}$ are two (arbitrary) Cauchy surfaces, $\Delta$ is a WIH horizon where an equivalence class of null normals $[\ell]$ has been fixed, and $\tau_\infty$ is an asymptotic time-like hypersurface.
For asymptotically flat spacetimes,
we are neglecting the boundary term in the asymptotic region, $S_{\text{SD}}^{\tau_\infty}$, that should be added in order to make the action differentiable at infinity. 
\cite{ABF, Thiemann:1993zq, CorichiReyes}. 
Here we are interested in the inner boundary, WIH, and in what follows we will not take into account the contributions from the asymptotic region, as they are well understood.

\subsection{Covariant approach: Symplectic structure}

The variation of the action on covariant phase space produces a boundary term 
\begin{equation}
\ed  S_{\text{SD}} (\delta)\vert_{\partial\mathcal{M}} := \delta S_{\text{SD}}\vert_{\partial\mathcal{M}} = -i\int_{\partial\mathcal{M}} \, \Sigma^{\;AB}\wedge\delta (\sdA_{\;AB}) \, .
\end{equation}
It can be shown that this term vanishes on $\Delta$, due to WIH boundary conditions \cite{CRV-2}, that is a necessary condition for the well posed variational principle. 

The second variation of the action is
\begin{equation}\label{IIVariation}
\ed^2  S_{\text{SD}} (\de_1,\de_2)=-2i\int_{\partial\mathcal{M}}\de_{[1}\S^{AB}\w\de_{2]} \,\leftidx{^+}{\A}{_{AB}}\, .
\end{equation}
Here the integral over $\tau_\infty$ (combined with a corresponding term that results from the variation of the boundary term that is needed at $\tau_\infty$) vanishes \cite{crv1}, but the contribution
from $\Delta$ does not. Nevertheless, on $\Delta$ the integrand in (\ref{IIVariation}) may be written in the form $\md j(\de_1,\de_2)$. This allows one to construct, by the standard procedure reviewed in section \ref{subsec:2.1}, a well defined presymplectic structure with a boundary term on $S_\Delta:=\Sigma\cap\Delta$, as in (\ref{CovSymplStructBT}). Let  us analyse in detail how this term comes about.

From the expansions of $\sigma^{\;AA'}$ and $\leftidx{^+}{\A}{_{AB}}$, given in (\ref{SL2CsolderingExpansion}) and (\ref{SDAexpansion}), one obtains
\begin{equation}
\Sigma^{AB}\wedge\leftidx{^+}{\A}{_{AB}}=2\,l\wedge m\wedge\bar{\U}+(m\wedge\bar{m}-l\wedge k)\wedge(\W+\V)
+2\,k\wedge\bar{m}\wedge\Y\,,
\end{equation}
The pullback to $\Delta$ leaves only one term in this expression
\begin{equation}
\Sigma^{AB}\wedge\leftidx{^+}{\A}{_{AB}}\WIHeq -i\, \leftidx{^2}{\epsilon} \wedge(\omega + V)  \,,
\end{equation}
so that
\begin{equation}
 \delta S_{\text{SD}}\vert_{\Delta} = -\int_{\Delta} \, \leftidx{^2}{\epsilon}\wedge \delta (\omega + V)\, .
\end{equation}

The WIH contribution to the second variation of the action is
\begin{equation}
\ed^2  S_{\text{SD}} (\de_1,\de_2)\vert_{\Delta}=-2\int_{\Delta} \, \delta_{[1}(\leftidx{^2}{\epsilon})\wedge \delta_{2]}(\omega +  V)
\, .\label{SecondVarWIH}
\end{equation}
Following \cite{AFK}, let us introduce two new fields, the real and imaginary potentials $\psi_R$ and $\psi_I$ respectively, such that
\begin{eqnarray}
\mathcal{L}_{\ell}{\psi_R} &=& \ell\cdot\md\psi_R\WIHeq \ell\cdot\omega\, ,\label{real_psi}\\
\mathcal{L}_{\ell}{\psi_I} &=& \ell\cdot\md\psi_I\WIHeq \ell\cdot V\, .\label{imaginary_psi}
\end{eqnarray}
These equations prescribe each $\psi_R$ and $\psi_I$ from the fields $\omega$ and $V$, up to a function $f$ on $\Delta$ such that $\mathcal{L}_\ell f=0$. Therefore, without further conditions on the definition of these fields, this ambiguity effectively introduces new degrees of freedom into the formalism. In order to avoid this,  again following \cite{AFK}, we may set $\psi_R$  to a fixed function on $S_{1\D}=\Sigma_1\cap\D$ for all points on $\Gamma_\text{cov}$\footnote{Recall that because of the way we construct the covariant phase space, all the points in $\Gamma_\text{cov}$ consist of the same differentiable manifold $\mathcal{M}$, and in particular share the same boundary, including $\Sigma_{1,2}$.}. The first order equation (\ref{real_psi}) together with initial conditions on $S_{1\D}$ completely determine $\psi_R$ from the (constant) field $\kappa_{(\ell)}=\ell\cdot\omega$ on each spacetime in the covariant phase space.
Notice that independently fixing  $\psi_R$  on $S_{1\D}$ to different values on each spacetime seems enough to eliminate or freeze the extra degrees of freedom introduced by $\psi_R$.
As we shall argue, the more stringent choice of fixing  $\psi_R$ at $S_{1\D}$ to the same function at each spacetime, prevents the `freedom' of choosing the initial value of $\psi_R$ at each spacetime from being interpreted as dynamical flow on covariant phase space.
Momentarily, we will not impose any analogous boundary conditions on $\psi_I$. Hence, a priori the definition of $\psi_I$ introduces additional degrees of freedom on $\Delta$. However, we shall see that these `boundary degrees of freedom' may be interpreted as $U(1)$ gauge for the transverse connection potential $V$, or may also consistently be frozen.

Taking into account that $\md\leftidx{^2}{\epsilon}\WIHeq 0$, and that
$\leftidx{^2}{\epsilon}\wedge\omega \WIHeq \leftidx{^2}{\epsilon}\wedge\md\psi_R$ and   
$\leftidx{^2}{\epsilon}\wedge V \WIHeq \leftidx{^2}{\epsilon}\wedge\md\psi_I$ (independently on any initial conditions on $\psi_R$ and $\psi_I$), we may rewrite (\ref{SecondVarWIH}) as
\begin{equation}
\ed^2  S_{\text{SD}} (\de_1,\de_2)\vert_{\Delta}=-2\int_{\Delta} \,\md\, \bigl( \delta_{[1}(\eps )\,\delta_{2]}\psi\bigr)
\, ,\label{SecondVarWIHTotalDeriv}
\end{equation}
where $\psi := \psi_R+\psi_I$. 
As noted in \cite{AFK}, the orientation of $S_\Delta$ is induced from $\Sigma$ (not from $\Delta$)\footnote{Due to this choice one can obtain the correct equations of motion on the WIH, as we will show in the following.}, so that
\begin {equation}
 \ed^2  S_{\text{SD}} (\de_1,\de_2)\vert_{\Delta} = - (\int_{S_{1\Delta }}-\int_{S_{2\Delta}})\bigl( -2 \delta_{[1}(\eps )\,\delta_{2]}\psi \bigr)\, .
\end {equation}
Consequently, the presymplectic structure may be defined as
\begin{equation}
\pmb{\omega}_{\text{SD}} (\de_1,\de_2)=-2i\int_{\Sigma}\de_{[1}\S^{AB}\w\de_{2]} \,\leftidx{^+}{\A}{_{AB}}
+ 2\int_{S_\D}\de_{[1}(\eps )\,\de_{2]} \psi \,. \label{pal_ss1}
\end{equation}

As already anticipated and further explained in the next section, the presence of the boundary term in (\ref{pal_ss1}) does not necessarily imply boundary degrees of freedom. This interpretation depends on the exact definition of the fields $\psi$, in particular, on the initial conditions we may or may not impose on $\psi_I$ (and $\psi_R$) and consequently on the nature of the variations $\delta_{1,2}\psi$.


\subsection{Covariant approach: Hamiltonian on $\Delta$}

The Hamiltonian $H_t$ is a conserved quantity that generates time translations in the covariant phase space. 
The contribution to the energy from a WIH 
was identified in \cite{AFK,cg} with the boundary term at $\Delta$ of this Hamiltonian. The Hamiltonian corresponds to  evolution vector fields $t^\mu$ that belong to the equivalence class $[\ell]$ defining a WIH. 
This equivalence class is the analog of constant multiples of a Killing vector field for weakly isolated horizons. In the cases of the Palatini and Holst actions, the 
first law of mechanics of non-rotating black holes has been deduced directly from the condition that the infinitesimal time translations $\delta_t$ define Hamiltonian vector fields on $\Gamma_\text{cov}$.

Following \cite{AFK}, let us consider a case when $t^\mu$ induces time evolution
on the covariant phase space, generated by a vector field $\de_t :=(\mathcal{L}_t \Sigma^{AB},\mathcal{L}_t \leftidx{^+}{\A}{_{AB}}  )$.
At infinity $t^\mu$ approaches a
time-translation Killing vector field of the asymptotically flat spacetime. On the 
other hand, on the non-rotating horizon, it should belong to the 
equivalence class $[\ell]$.
This makes sense since all points on $\Gamma_\text{cov}$ share the same asymptotic structure and WIH structure $[\ell]$ at the corresponding boundaries. However, unlike the case at $\tau_\infty$ where there is a unique asymptotic flat metric and thus a common time symmetry direction, at $\Delta$, $t^\mu$ may (and has to) be identified with different elements of the equivalence class for different points in $\Gamma_\text{cov}$.
The vector field  $\de_t$ will be Hamiltonian, or equivalently represent a phase space symmetry 
$\La_{\de_t}\pmb{\omega}_{\text{SD}}=0$,  iff the one-form
\be
X_t (\de )=\pmb{\omega}_{\text{SD}} (\de ,\de_t )\, , \label{oneformFromOmegat}
\ee
is closed, and the Hamiltonian $H_t$ is defined as 
\be
\de H_t :=X_t (\de )= \de E^t_{\text{ADM}} - \de E^t_\Delta\, ,
\ee 
where $E^t_{\text{ADM}}$ is the ADM energy (corresponding to the boundary term at $\tau_\infty$ obtained when one includes the term $S_{\text{SD}}^{\tau_\infty}$) and $E^t_\Delta$ (the boundary term at $\Delta$) can be interpreted as horizon energy.

The symplectic structure $\pmb{\omega}_{\text{SD}} (\de ,\de_t )$ in (\ref{oneformFromOmegat}) has two contributions, 
\begin{equation}
\pmb{\omega}_{\text{SD}} (\de,\de_t)=-i\int_{\Sigma}\bigl(\de\S^{AB}\w\La_t (\leftidx{^+}{\A}{_{AB}})-\La_t\S^{AB}\w\de (\leftidx{^+}{\A}{_{AB}})\bigr)
- \int_{S_\D}\bigl(\de\psi\,\de_t (\eps ) - \de_t\psi\,\de (\eps )\bigr)\, 
.\label{pal_ss2}
\end{equation}

To evaluate the second term in this expression we must first define or determine the values of the boundary variations $\left(\de_t (\eps),\de_t\psi\right)$ from the `bulk variations' or vector field $\de_t =(\mathcal{L}_t \Sigma^{AB},\mathcal{L}_t \leftidx{^+}{\A}{_{AB}}  )$. Since the metric on $\Delta$ and thus $\eps$ is determined directly from the basic fields $(\Sigma,\leftidx{^+}{\A})$, one must have $\de_t(\eps)=\mathcal{L}_t\eps$. And as we have reviewed, WIH boundary conditions imply $\La_t \eps\WIHeq 0$. In contrast, determining $\de_t\psi=\de_t\psi_R+\de_t\psi_I$ is more subtle. As already pointed out in  \cite{AFK,cg,ABL}, $\de_t\psi_R$ cannot be defined as $\La_t\psi_R$. This  comes precisely
because one has fixed the same `initial value' of  $\psi_R$  on $S_{1\D}=\Sigma_1\cap\Delta$ for all points on $\Gamma_\text{cov}$. Effectively one has set for all possible variations on phase space $\delta\psi_R|_{S_{1\D}}=0$. In particular, for any dynamical vector field $\delta_t$ one must have $\delta_t\psi_R|_{S_{1\D}}=0$, which together with the WIH boundary conditions implies $\de_t\psi_R \WIHeq 0$. We will explain this important point in some detail. By definition, $t^\mu$ is a live vector field, it changes from point to point in the covariant phase space. However,  on $\Delta$, the vector fields corresponding to two arbitrary points on $\Gamma_\text{cov}$ satisfy $t'^\mu=\tilde{c} \,t^\mu$. For infinitesimally close points, the potential $\psi_R'=\psi_R+\delta_t\psi_R$ also satisfies $\La_{t'}\psi_R '=t'\cdot\omega'$. Since $\omega'=\omega+\La_t\omega \WIHeq\omega$, it follows that $\La_t\psi_R '=\k_{(t)}$. Therefore
$
\La_t\,(\de_t\psi_R )=\La_t\,(\psi_R '-\psi_R)\WIHeq 0\,.
$
That is, the variation $\delta_t\psi_R$ will be determined by its value at the  `initial slice' or sphere $S_{1\Delta}$.
As we have defined $\psi_R$ such that   $\de_t\psi_R = 0$ on $S_{1\Delta}$, this shows that on the whole WIH
\be  \label{infinitesimalGeneratorofTTphiR}
\de_t\psi_R\WIHeq 0\, .  
\ee
Note that the choice of individually fixing the value of $\psi_R$ at $S_{1\D}$ in each spacetime, does eliminate the freedom introduced by $\psi_R$ in each spacetime (namely, an arbitrary function $f$ on $\Delta$  such that $\mathcal{L}_\ell f=0$, or equivalently, an arbitrary function on the sphere  $S_{1\D}$).
However, one still has the freedom to choose this initial value at each spacetime, i.e. a freedom on how one defines or fixes the value of $\psi_R$ at each spacetime, and it is this assignment that ultimately determines the possible values of variations $\de\psi_R$ on phase space. If one does not assign a common initial value on $S_{1\D}$ for all spacetimes then $\delta\psi_R\neq 0$, which in general would also require $\delta_t\psi_R\neq 0$. 

On the other hand,
the definition $\delta_t\psi_I=\La_t \psi_I$ is consistent with the WIH boundary conditions. Indeed, since $\La_\ell V= -2i\, \md (\text{Im}\,\epsilon_{\text{NP}})$ \cite{cg,CRV-1}, with $\epsilon_{\text{NP}}$
a  Newman-Penrose spin coefficient,  it follows that 
\[
\La_{t'}\psi_I'=t'\cdot V'=t'\cdot (V+\La_{t}V )=
\La_{t'}\psi_I + c\,t'\cdot\md (\epsilon_{\text{NP}}-\bar{\epsilon}_{\text{NP}})=
\La_{t'}[\psi_I+c (\epsilon_{\text{NP}}-\bar{\epsilon}_{\text{NP}}) ]\, ,
\]
where $t^\mu=c\, \ell^\mu$. Since, 
$c(\epsilon_{\text{NP}}-\bar{\epsilon}_{\text{NP}}) = t\cdot V=\La_t \psi_I$, it follows that
\[
\La_t [\de_t\psi_I - \La_t \psi_I]=0 \, .
\]
That is, $\de_t\psi_I = \La_t \psi_I$ modulo a function on $\Delta$ whose Lie derivative along $t^\mu$ vanishes. This is precisely the apparent ambiguity introduced by $\psi_I$, and it is consistent with the fact that $t\cdot V$ is only dynamically determined up to $U(1)$ gauge.
In fact, $\psi_I$ exactly encodes the degrees of freedom of $t\cdot V$  (i.e. degrees of freedom of the transverse connection potential $V$ along $t^\mu$). The apparent additional freedom introduced by the definition of $\psi_I$, namely, the ambiguity of adding a function $\psi_I-i\theta$ such that $\mathcal{L}_t\theta=t\cdot\md\theta=0$, can be reinterpreted as special (`time independent') U(1) gauge: $V\to V+i\md\theta$.
In analogy with $\psi_R$, one may now eliminate this ambiguity or `redundant' extra degrees of freedom, by  fixing the initial value of $\psi_I$  on $S_{1\D}$ at each point in $\Gamma_\text{cov}$. However, one must do so consistently, i.e. in such a way that the general variations $\delta\psi_I$ in $\Gamma_\text{cov}$ allow $\de_t\psi_I=\La_t \psi_I$, and guarantee that no degrees of freedom from $V$ are suppressed in the dynamical flow generated by $\delta_t$ in covariant phase space.

If we assume that $\de_t\psi_I=\La_t \psi_I$ on 
$S_{1\D}$, it follows that on $\Delta$,
\be
\de_t\psi_I\WIHeq \La_t\psi_I=t\cdot V\, .
\ee
As a result, the corresponding variation of the potential $\psi =\psi_R+\psi_I$ is
\be\label{delta_psi}
\de_t\psi = t\cdot V\, .
\ee

With these results at hand, we may now go back to evaluate the symplectic structure in (\ref{oneformFromOmegat}).
As shown in \cite{AFK}, in the case of the Palatini action, the bulk term in $\pmb{\omega}_{\text{P}} (\de ,\de_t )$ is an integral of a total derivative. Similarly, in this case it turns out that the bulk term in $\pmb{\omega}_{\text{SD}} (\de ,\de_t )$ reduces to a boundary term, such that 
\be \label{cheatingEquation}
 -\de E_\Delta^t :=\pmb{\omega}_{\text{SD}} (\de ,\de_t )|_{S_\Delta}= \int_{S_\Delta} [-i\, (t\cdot\leftidx{^+}{\A}{_{AB}} )\,\de\S^{AB} +
(t\cdot V)\, \de (\eps) ]\, ,
\ee
where the first term represents the bulk contribution, and the second one is the boundary term in (\ref{pal_ss2}). Since, $(t\cdot\leftidx{^+}{\A}{_{AB}} )\,\de\S^{AB}\WIHeq -i\, t\cdot(\omega + V)\, \de(\eps)$ it follows that
\be
\de E_\Delta^t= \k_{(t )}\de a_\D\, ,
\ee
where $a_\D$ is the area of the horizon. This is the same result as in Palatini case \cite{AFK}. The formula implies $\k_{(t )}$ has to depend only on the horizon area and gives the first law of black hole mechanics as a consistency requirement for the evolution vector field $\delta_t$ defined by $t^\mu$ to be Hamiltonian. Selecting a permissible vector field is equivalent to prescribing a relation $\k_{(t )}(a_\D)$. To fix this ambiguity and define the horizon mass, one imposes 
\be
\k_{(t )}=\frac{1}{2R_\Delta}\,,    \label{staticKappaAreaRelation}
\ee
with $a_\Delta =4\pi R_\Delta^2$. This guarantees that for the static solution $t^\mu$ agrees  with the Killing vector field on $\Delta$. The horizon mass is
\be
M_\Delta = 2\,\k_{(t )} a_\D\, .  \label{HorizonMassDef}
\ee

At this point, a further explanation of our choices is in order. Notice first that the introduction of the potentials $\psi_R$ 
and $\psi_I$ is indispensable to show we have a conserved symplectic current and a well defined symplectic structure. Nevertheless, there are two possible interpretations to the choices we have just made. On the one hand, one can take the viewpoint that there is only one covariant phase space $\Gamma_\text{cov}$ (consisting of solutions to the equations of motion that satisfy the appropriate boundary conditions).  This seems to be the point of view taken in \cite{AFK, ABL} and the one we have followed so far.
As we have argued, if as part of their definition, one fixes the initial value of $\psi_R$ on all of $\Gamma_\text{cov}$ and the initial value of $\psi_I$ consistently on each spacetime, by the nature of the WIH boundary conditions, one effectively is prescribing their value on each spacetime. This definition guarantees no additional (dynamical) degrees of freedom are introduced (or original ones suppressed). 
Despite the form (\ref{pal_ss1}) of $\omega_\text{SD}$, there are no boundary degrees of freedom in $\Gamma_\text{cov}$. As we have seen, the value of $\psi=\psi_R+\psi_I$ is completely determined by  $\omega$ and $V$, which in turn are determined by the value of $^+\mathcal{A}_{IJ}$ in the bulk. The boundary variations $(\de (\eps ),\de \psi)$ in (\ref{pal_ss1}) are not independent, they are prescribed by the variations in the bulk  $(\de \Sigma^{AB},\de \leftidx{^+}{\A}{_{AB}} )$.
The exact value of the these boundary variations certainly depends on our choice of initial conditions for the potentials, but the formalism seems to be insensitive to this choice.

On the other hand, our analysis already shows that one can introduce  $\psi_I$  as  boundary degree of freedom, by not fixing its initial value  at all, in a way consistent with the true dynamics in the bulk. Since $\de \psi_R$ and $\de \psi_I$ appear explicitly in the symplectic structure regardless of any `initial' conditions, one can consider equations (\ref{real_psi}) and (\ref{imaginary_psi}), without any further restrictions, as the definitions for $\psi_R$ and $\psi_I$. Consequently, these are now new boundary degrees of freedom on $\Delta$. Thus from this perspective, $\omega_\text{SD}$ in (\ref{pal_ss1}) represents a  symplectic structure on an `extended' phase space $\widetilde{\Gamma}_\text{ext}$ whose points or spacetimes carry these additional degrees of freedom, and where the variations $(\de (\eps ),\de \psi)$ are truly independent from bulk variations.
From this point of view, the dynamical evolution of the independent fields $(\eps , \psi)$ must now be defined as
\be
\de_t(\eps):=\La_t\eps \quad \text{and} \quad \de_t\psi_I:=\La_t\psi_I\,,  \label{trueFlowEPsiI}
\ee
 for consistency with the bulk fields, and as 
 \be
 \delta_t\psi_R:=0\,,
 \ee
in order to freeze the extra degrees of freedom from $\psi_R$ so that they remain 'kinematical' and not 'dynamical'. These definitions ensure the Hamiltonian flow in $\widetilde{\Gamma}_\text{ext}$ matches the Hamiltonian flow in $\Gamma_\text{cov}$.

The latter point of view is the one more readily adapted to the canonical formalism. In fact, our analysis suggests at least three different possible consistent extensions of $\Gamma_\text{cov}$. 
\begin{description}
\item [Covariant extension I:] 
Consider both $\psi_R$ and $\psi_I$ as boundary degrees of freedom, by not imposing any initial conditions in their definition. Notice again there are no consistency restrictions in the definition of the dynamics of $\psi_R$. One may get  a formulation with consistent dynamics for different values of $\delta_t\psi_R$. However $\delta_t\psi_R\neq 0$, results in additional terms in (\ref{cheatingEquation}).
In particular, for the natural (unrestricted) dynamical flow $\delta_t\psi=t\cdot(\omega+V)$, we get consistent dynamics but the `wrong' value $\de E_\Delta^t=0$ for the energy since the Hamiltonian flow does not match the original flow.
\item [Covariant extension II:] 
Consider  $\psi_I$ as boundary degree of freedom, by  imposing  initial conditions on $\psi_R$ only. As we have already remarked, in this case we get consistent dynamics that reproduce the original flow and hence give the correct value for the energy $\de E_\Delta^t= \k_{(t )}\,\de a_\D$.
This case is of significance because it shows we can consistently reinterpret the (bulk) degrees of freedom as independent boundary degrees of freedom.
\item [Covariant extension III:] Consider  $\psi_R$ as boundary degree of freedom, by  imposing  initial conditions on $\psi_I$ only.
As in case I, here too, one obtains a well defined formulation also with the unrestricted flow $\delta_t\psi=t\cdot(\omega+V)$, but with vanishing energy. Also, notice that if one does not fix the values of $\psi_I$ on $S_{1\D}$ consistently, or simply sets $\de_t\psi_I=0$ (ignoring consistency condition (\ref{trueFlowEPsiI})), one also gets a consistent formulation but, as expected, with the wrong expression for the energy too. Indeed, the last term in (\ref{cheatingEquation}) would vanish and we would get $\de E_\Delta^t=\int_{S_\Delta} (\k_{(t )}+t\cdot V)\de \eps$. 
\end{description}
Finally, we remark once more that the covariant symplectic structure, equation (\ref{pal_ss1}), 'looks the same' in all cases, but it is obviously distinct in each of these extensions and in the original covariant phase space.

Let us now analyse the same theory within the canonical approach, and see how one can obtain corresponding symplectic structures  and  Hamiltonians in corresponding extensions of the canonical phase space.


\subsection{Canonical approach: Symplectic structures}
 
For a 3+1 decomposition and canonical formulation of a covariant action, one postulates a time function $t:\mathcal{M}\to \mathbb{R}$, where the region of the spacetime is foliated as $\mathcal{M}=I\times\Sigma_t$,    where $\Sigma_t$ are level curves $t={\text{const}}$.  The foliation is chosen to be \textit{compatible} with the isolated horizon structure, in the sense that it induces a foliation of $\Delta$ by two-spheres $S_t=\Sigma_t\cap\Delta$ and the null geodesic generators of $\Delta$ parameterized by $t$ belong to the equivalence class $[\ell]$.
Additionally, one needs to choose an \textit{evolution} vector field $t^\mu$ such that $t^\mu\nabla_\mu t=1$, 
that is generically time-like on the bulk and becomes null as one approaches  $\Delta$. For non-rotating horizons it is also chosen to belong to the equivalence class $[\ell]$ on $\Delta$ but, unlike the covariant case, there is no need for live vector fields so we can single out a representative of the equivalence class for all points in $\Gamma_\text{can}$ by fixing the lapse function $N$. 

Decomposition of the spacetime metric splits the ten independent components of $g_{\mu\nu}$ into the six independent components of the Euclidean spatial metric $q_{ab}$, the lapse function $N$ and the  shift vector $N^\mu$, such that $t^\mu=Nn^\mu+N^\mu$, where  $n^\mu$ denotes the future directed unit normal to the foliation.

The canonical action is obtained starting from the covariant action (\ref{SDactionVer2}), performing  3+1 decomposition of the fields, as in \cite{ABF,CRV-2}, and taking into account that \cite{Espanolitos,ABF}
\begin{equation}
 S_{\text{SD}}=\int_{\mathcal{M}}{\pmb{L}}_{\text{SD}}=-i\,\int_I\md t\, \int_\Sigma \,
\bigl[ t\cdot (\Sigma^{AB}\wedge \sdF_{AB} )\bigr] \, . 
\end{equation}

Spacetime fields are split into tangential (spatial) and transverse components with respect to this foliation.
The four-dimensional  $SL(2,\mathbb{C})$ soldering form on $\mathcal{M}$, $\sigma^\nu_{\;AA'}$, induces a three-dimensional   $SU(2)$ soldering form on $\Sigma$, $\sigma^\mu_{\;A}\,^B$, in the following way  \cite{ACKclassical}
\begin{equation} \label{SU2solderingForm}
\sigma^\mu_{\;A}\,^B:=-i\sqrt{2}\,q^\mu_{\,\nu}\,\sigma^\nu_{\;AA'}\,n^{A'B}\, ,
\end{equation}
where $q^\mu_{\,\nu}:= \delta^\mu_\nu + n^\mu n_\nu$ is a projector on $\Sigma$ and $n^{A'B}=n^\mu\sigma_\mu^{\;A'B}$ is the spinorial representation of $n^\mu$.

From this one obtains the canonical action (see, for example, \cite{ABF,JacobsonSmolin}) 
\begin{equation}
S_{\text{SD}}^{\text{can}} 
=\int_I\md t\, \big\{\int_\Sigma\md^3y\bigl( -i\, \Sigma^{AB}\wedge {\mathcal{L}}_{t}A_{AB}\bigr)
-H[N]  -C[N^a]  + G[t^\mu\,\sdA_\mu^{\;AB}]\,\big\} + S_{\text{SD}}^{\Delta} \, , \label{CanAct1}
\end{equation}
where $\Sigma^{AB}$ and $A_{AB}$ are the spatial components of the corresponding four dimensional fields, the forms pulled back to $\Sigma$.

The other bulk terms in $S_{\text{SD}}^{\text{can}}$ are (smeared) first class constraints, the Hamiltonian constraint $H[N]$,
the vector constraint $C[N^a]$ and 
the Gauss constraint, $G[\Lambda^{AB}]$ (their explicit form is given, for example, in \cite{CRV-2}).
The boundary term $S_{\text{SD}}^{\Delta}$ is obtained after integration by parts.

The kinetic term in the canonical action (\ref{CanAct1}) determines the bulk symplectic structure
\begin{equation}
\Omega^{\text{Bulk}}_{\text{SD}}(\delta_1,\delta_2)=
-2i\, \int_\Sigma \delta_{[1}\Sigma_{AB} \wedge \delta_{2]}A^{AB}\, .
\end{equation}

As shown in \cite{CRV-2}, the WIH boundary term plays a crucial role in the canonical analysis.
It is of the form
\be
S_{\text{SD}}^{\Delta}=-\int_I \md t \oint_{S_\Delta} t\cdot (\omega +  V )\,\leftidx{^2}{\epsilon} \, .\label{GaugeFixedSDboundary1}
\ee
Following the same path as in the covariant case, we may introduce $\psi =\psi_R + \psi_I$, as in (\ref{real_psi}) and (\ref{imaginary_psi}).
The fact that one can get consistent formulations by independently considering $\psi_R$ or  $\psi_I$  as extra degrees of freedom in the covariant formalism, suggests corresponding extensions of the canonical phase space.
In the canonical formalism we have an additional ambiguity, or two possible ways to proceed. Following the covariant procedure one may introduce the potentials along with `initial conditions'  for either or both of them at some time $t_0$, and rewrite the boundary term as
 \begin{equation} \label{CanSympl1}
 S_{\text{SD}}^{\Delta}=-\int_I \md t \oint_{S_\Delta} (\mathcal{L}_{t}\, \psi )\,\leftidx{^2}{\epsilon} \, .
\end{equation}
This term gives a canonical symplectic structure of the form
\be \label{SymplStructSD1}
\Omega_{\text{SD}}(\de_1,\de_2) =-2i\, \int_\Sigma \delta_{[1}\Sigma_{AB} \wedge \delta_{2]}A^{AB} -  2\oint_{S_\Delta} \delta_{[1} (\leftidx{^2}{\epsilon}) \delta_{2]} \psi\, .
\ee
regardless of whether we consider  $\psi_R$ or  $\psi_I$ as boundary degrees of freedom or not, just like in the covariant case. Correspondingly, one has a candidate Hamiltonian with no boundary terms. As we shall argue, this procedure gives a consistent canonical formulation with no boundary degrees of freedom (if one fixes initial conditions for both $\psi_R$ and $\psi_I$), and three consistent canonical extensions corresponding to the covariant extensions analysed before. The corresponding canonical symplectic structures (\ref{SymplStructSD1}) will be $\bar{\Pi}$ - equivalent, as defined in \cite{CRV-3}, with the covariant symplectic structures (\ref{pal_ss1}). 

Additionally, in the canonical formalism one may also introduce $\psi_R$ or  $\psi_I$  as extra degrees of freedom independently, and reinterpret each term in (\ref{GaugeFixedSDboundary1}) as  kinetic or Hamiltonian accordingly. This is the point of view taken in \cite{CRV-2} which we will also review and refine here in light of the covariant analysis.
First, we already have the case where one may consider the ambiguities brought about by the definition of both real and imaginary potentials, as boundary degrees of freedom. In this instance, one rewrites the boundary term as (\ref{CanSympl1}), and consistently interprets it  as a boundary kinetic term. This term contributes to a boundary term in the canonical symplectic structure which takes the form (\ref{SymplStructSD1}), with boundary variations $\delta\psi=\delta\psi_R+\delta\psi_I$ independent from bulk variations.

Similarly, one can only introduce $\psi_I$ as boundary degrees of freedom, and rewrite boundary term (\ref{GaugeFixedSDboundary1}) as
\begin{equation}\label{SboundarySP}
 S_{\text{SD,2}}^{\Delta} =-\int \md t\oint_{S_\Delta} \, (\kappa_{(t)}\,  +  \mathcal{L}_t\psi_I ) \,\leftidx{^2}{\epsilon} \, .
\end{equation}
In this instance, one term contributes to the boundary term in the (candidate) canonical Hamiltonian and the other one  to the boundary symplectic structure, making our interpretation consistent again. The symplectic structure now reads
\be\label{SymplStructSD2}
\Omega_{\text{SD,2}}(\de_1,\de_2) =-2i\, \int_\Sigma \delta_{[1}\Sigma_{AB} \wedge \delta_{2]}A^{AB} - 2\oint_{S_\Delta} \delta_{[1} (\leftidx{^2}{\epsilon}) \delta_{2]} \psi_I\, .
\ee

One can also take the opposite viewpoint to the previous one and choose $(\psi_R,\leftidx{^2}{\epsilon})$ as boundary degrees of freedom, while interpreting the term with $t\cdot V$ as contributing to the Hamiltonian. In that case
\be\label{SymplStructSD3}
\Omega_{\text{SD,3}}(\de_1,\de_2) =-2i\, \int_\Sigma \delta_{[1}\Sigma_{AB} \wedge \delta_{2]}A^{AB} - 2\oint_{S_\Delta} \delta_{[1} (\leftidx{^2}{\epsilon}) \delta_{2]} \psi_R\, .
\ee

The last possibility is 
not to extend the canonical phase space,  leave the boundary term in the canonical action in its original form (\ref{GaugeFixedSDboundary1}) and interpret it as a term contributing to the Hamiltonian.
In this case there are no boundary contributions to the symplectic structure
\be\label{SymplStructSD4}
\Omega_{\text{SD,4}}(\de_1,\de_2) =-2i\, \int_\Sigma \delta_{[1}\Sigma_{AB} \wedge \delta_{2]}A^{AB} \, ,
\ee
and no boundary degrees of freedom, 
As a consequence the variations of the allowed functionals cannot have a boundary term, as in the Regge-Teitelboim approach \cite{Regge&Teitelboim}.

In all of the cases,  relation (\ref{HVF1}) must be satisfied, and will prescribe the boundary equations of motion. Consistency can lead to the need to introduce a counterterm in the corresponding canonical Hamiltonian, or to further restrict boundary conditions. We will review these possibilities in what follows.

\subsection{Canonical approach: Hamiltonians on $\Delta$ }

In the canonical framework the equations of motions are given by 
\begin{equation}
\ed H_{\text{SD}} (Y) =\Omega (Y ,X_H)\, , \label{HamEqs}
\end{equation}
where $X_H$ is the corresponding Hamiltonian vector field.
The bulk part of (\ref{HamEqs}) gives the equations of motion in the bulk, while
the boundary term  defines the boundary components of the Hamiltonian vector field. Here, we are specially interested in the boundary contributions in (\ref{HamEqs}), that depend on the symplectic structure. 
Let us analyse and extend the results of \cite{CRV-2}, separating into four possible cases corresponding to the specific additional boundary degrees of freedom introduced in the formalism and the  different interpretations of the boundary term (\ref{CanAct1}) in each case.
\subsubsection{{\bf{ Canonical extension with both $\psi_I$ and $\psi_R$ as boundary degrees of freedom}}}
If we choose to write the boundary term in (\ref{CanAct1}) as  (\ref{CanSympl1})
without any restrictions on the potentials, we are introducing boundary degrees of freedom and enlarging the original canonical phase space. In this case we have to interpret the boundary term as a kinetic term, and the canonical Hamiltonian is just the linear combination of constraints,
\begin{equation}\label{Ham1}
  H_{\text{SD,1}}= H[N]  +C[N^a]  - G[t^\mu\,\sdA_\mu^{\;AB}]\, ,
\end{equation}
such that, as shown in \cite{CRV-2},
\be\label{VarH1}
\delta H_{\text{SD,1}}|_{S_\Delta} = -\oint_{S_\Delta} t\cdot (\omega +V)\delta \,\leftidx{^2}{\epsilon}\, .
\ee
On the other hand
\be
\Omega_{\text{SD,1}} (\de ,\de_t)\vert_{S_\Delta} = -\oint_{S_\Delta} (\de_t\psi\,\de\,\eps - \de\psi\,\de_t\eps )\, .
\ee
The comparison of  the two expressions allows one to read off the boundary components $(\delta_t \psi,\delta_t\eps)$ of the Hamiltonian vector field:
\be\label{Var1}
\delta H_{\text{SD,1}}|_{S_\Delta}=\Omega_{\text{SD,1}} (\de ,\de_t)\vert_{S_\Delta}\ \ \Rightarrow\ \ \de_t\psi=t\cdot (\omega +V)\ \ \text{and}\ \ \de_t\eps =0\, ,
\ee
which give consistent Hamiltonian evolution equations for the boundary degrees of freedom $(\psi,\eps)$ on $S_\Delta$:
 \be \label{HamiltonEoMC1}
\La_t\psi= \de_t\psi  = t\cdot (\omega +V) \qquad \text{and} \qquad    \La_t\eps= \de_t\eps =0\,.
 \ee
 
 We can conclude that the Hamiltonian $H_{\text{SD,1}}$ is well defined,
 and gives a consistent evolution, but it vanishes on the constraint surface. In this extended canonical phase space one cannot obtain the correct expression for the energy of $\Delta$. 
This is consistent with the zero energy value given by the natural evolution of \textit{Covariant extension I} in the covariant treatment, where one does not fix the value of $\psi$ on each spacetime.

Accordingly, one must now note that,
if one wants  the spacetimes defined by the points of $\Gamma_\text{can}$ to ``match" the spacetimes or points in $\Gamma_\text{cov}$, then equations of motion (\ref{HamiltonEoMC1}) defined by this Hamiltonian are inconsistent. 
The natural Hamiltonian flow in this extended phase space does not match the original dynamical flow.
More precisely, one should require that all the points in the constraint surface $\bar{\Gamma}_\text{can}$ of the canonical phase space be in the image of the canonical projection $\bar{\Pi}:\Gamma_\text{cov}\to\bar{\Gamma}_\text{can}$ assigning to each solution in $\Gamma_\text{cov}$ its `instantaneous' value at some initial hypersurface $\Sigma_1$.
The projection or push forward of the infinitesimal generator of time evolution $\delta_t$ on covariant phase space should be in the direction of the Hamiltonian flow for time evolution in canonical phase space. In particular the component $\delta_t\psi_R\WIHeq 0$, equation (\ref{infinitesimalGeneratorofTTphiR}), must project to a Hamiltonian vector field component with $\delta_t\psi_R=0$ and therefore $\de_t\psi =t\cdot V$.
This is the Hamiltonian vector field that `freezes' the extra degrees of freedom introduced by $\psi_R$.  
The symplectic structure, when  acting on $(\delta\, , \delta_t)$ simplifies
\be
\Omega_{\text{SD,1}} (\de ,\de_t)\vert_{S_\Delta} = -\oint_{S_\Delta} \left((t\cdot V)\,\de\eps - \delta\psi\,\delta_t\eps\right) \, .
\ee
Relation (\ref{HamEqs}) consistently sets $\delta_t\eps=0$ again, but now one needs to add a counterterm to the Hamiltonian, $H_{\text{CT,1}}$, such that
\be
\de (H_{\text{SD,1}}+H_{\text{CT,1}})|_{S_\Delta}=-\oint_{S_\Delta} (t\cdot V)\,\de\eps \, ,
\ee
or equivalently,
\be
\de H_{\text{CT,1}}=\oint_{S_\Delta} (t\cdot \omega)\,\de\eps = \kappa_{(t)}\de a_\Delta\, , \label{CT1variation}
\ee
where $a_\Delta=\oint_{S_\Delta}\eps$ is the horizon area.

Following arguments already outlined in \cite{CRV-2}, the last equation implies that the counter term, and consequently the surface gravity $\kappa_{(t)}$, must be functions of the horizon area $a_\Delta$ only: $H_{\text{CT,1}}=\oint_{S_\Delta} F(a_\Delta)\,\eps=F a_\Delta$. Condition (\ref{CT1variation}) now gives a simple differential equation
\[
\frac{\md (Fa_\Delta)}{\md a_\Delta} =\kappa_{(t)}\,,
\]
which determines the counter term (up to a constant) and gives a consistent Hamiltonian evolution for every functional relation $\kappa_{(t)}(a_\Delta)$. This is the same ambiguity as in the covariant case for permissible evolution vector fields $\delta_t$. In particular, if one chooses relation (\ref{staticKappaAreaRelation}) so that
\be
\delta\kappa_{(t)}=-(\kappa_{(t)}/2a_\Delta)\delta a_\Delta\,,   \label{area_kappa}
\ee
 it turns out that $H_{\text{CT,1}}= 2\oint_{S_\Delta} \kappa_{(t)}\eps=2\kappa_{(t)}a_\Delta$, so that the Hamiltonian that satisfies (\ref{HamEqs}) takes the form
\be\label{CT1}
{\tilde{H}}_{\text{SD,1}}= H[N]  +C[N^a]  - G[t^\mu\,\sdA_\mu^{\;AB}] +2\oint_{S_\Delta} \kappa_{(t)}\eps\, .
\ee
The boundary term corresponds to the Smarr expression (\ref{HorizonMassDef}) for the mass of a non-rotating black hole \cite{Smarr}.  

In summary, this canonical extended phase space matches the 
\textit{Covariant extension I}, with the same (extra) boundary degrees of freedom $\psi$. The corresponding pre-symplectic structures are $\bar{\Pi}$-equivalent, as defined in \cite{CRV-3} 
, and consequently the correct dynamics can be consistently implemented, determining the mass of the WIH uniquely.

\subsubsection{\bf{Canonical extensions with $\Psi_I$ as  boundary degree of freedom}}
If we now rewrite the boundary term $S_{\text{SD}}^\Delta$ as in (\ref{SboundarySP}), then  
the canonical Hamiltonian acquires a boundary term,
\begin{equation}\label{Ham2}
  H_{\text{SD,2}}= H_{\text{SD,1}} + \oint_{S_\Delta} \, \kappa_{(t)}\,\leftidx{^2}{\epsilon} \, ,
\end{equation}
and we can see from (\ref{SymplStructSD2}) that $(\psi_I, \eps)$ are canonical variables on $\Delta$. 
As we have argued in the covariant analysis, this particular `extension' of the canonical phase should be consistent with the original theory, since one is not adding new real degrees of freedom.

In order to  calculate the boundary contribution to (\ref{HamEqs}), let us first analyse the variation of the Hamitonian, that is given by
\begin{equation}
\delta H_{\text{SD,2}}\vert_{S_\Delta} 
 =\oint_{S_\Delta}\, \bigl[ \delta\kappa_{(t)}\,\eps  -  (t\cdot V)\, \delta\, \eps \bigr]\, .\label{VarHamBound}
\end{equation}
From (\ref{SymplStructSD2}) one obtains
\be\label{SymplStructSD2Var}
\Omega_{\text{SD,2}}(\de ,\de_t)|_{S_\Delta} =- \oint_{S_\Delta} \de_t\psi_I\,\de\,\eps - \de\psi_I\,\de_t\eps\, ,
\ee
and we can see that the Hamiltonian $H_{\text{SD,2}}$ is not differentiable, since $\delta H_{\text{SD,2}}\vert_{S_\Delta}\neq\Omega_{\text{SD,2}}(\de ,\de_t)|_{S_\Delta}$. 
One needs a counter term to eliminate the first term in (\ref{VarHamBound})
\[
\de H_{\text{CT,2}}=-\oint_{S_\Delta}\, \de\kappa_{(t)}\,\eps\,.
\]
As shown in \cite{CRV-2}, analogously to the analysis in the previous case, this equation also implies $\kappa_{(t)}$ and the counter term are functions of area $a_\Delta$, and  the differential equation
\[
\frac{\md (Fa_\Delta)}{\md a_\Delta} =-a_\Delta\frac{\md \kappa_{(t)}}{\md a_\Delta}\,,
\]
now determines the counter term and energy for each functional relation $\kappa_{(t)}(a_\Delta)$.
If one chooses again (\ref{staticKappaAreaRelation}) with (\ref{area_kappa}), one gets a counterterm $H_{\text{CT,2}}=\oint_{S_\Delta}\, \kappa_{(t)}\,\leftidx{^2}{\epsilon}\,$, so
 a well defined Hamiltonian ${\tilde{H}}_{\text{SD,2}} := H_{\text{SD,2}}+H_{\text{CT,2}}$ is given by
 \begin{equation}\label{TotalHamCT}
 \tilde{H}_{SD,2}=\int_\Sigma\md^3y\, \big\{ 
H[N] + C[N^a]  - G[t^\mu\,\sdA_\mu^{\;AB}]\,\big\} + 2\oint_{S_\Delta} \, \kappa_{(t)}\,\leftidx{^2}{\epsilon} \, ,
\end{equation}
just like in the previous case, leading to the correct expression for the mass of a non-rotating black hole. 


Let us analyse now the canonical description of the evolution generated by ${\tilde{H}}_{\text{SD,2}}$ in the phase space. Since $\delta{\tilde{H}}_{\text{SD,2}}|_{S_\Delta} = \Omega_{\text{SD,2}}(\de ,\de_t)|_{S_\Delta}$ it follows that
\begin{eqnarray}
 \de_t\eps = 0 \ \ \ &\Rightarrow & \ \ \ \mathcal{L}_t\leftidx{^2}{\epsilon}=\de_t\eps=0\, ,\\
 \de_t\psi_I = t\cdot V\ \ \ &\Rightarrow & \ \ \  \mathcal{L}_t\psi_I=\de_t\psi_I = t\cdot V\, ,
\end{eqnarray}
which are consistent boundary equations, as in Case I. The boundary variations correspond to the ones in the covariant phase space, but without the need to modify the HVF at the horizon. As expected from the covariant analysis, reinterpreting some  of the degrees of freedom of the connection potential $V$ as boundary degrees of freedom, allows for a consistent formulation and to arrive at the expression for the energy and the first law.

Notice that the canonical extension above does not correspond to \textit{Covariant extension II}. As we already mentioned, in the covariant analysis one also has to introduce $\psi_R$ and treat it as a fixed function by setting initial conditions at some time $t_0$. The canonical structure is (\ref{SymplStructSD1}), with only the $\de \psi_I$ variations treated as independent. Calculations reduce to those of \textit{Canonical extension I} with $\de_t\psi_R=0$, giving thus also a consistent formulation.

\subsubsection{\bf{Canonical extensions with $\psi_R$ as boundary degree of freedom}}

Here we shall take the opposite viewpoint to the previous one, we shall consider $(\psi_R,\leftidx{^2}{\epsilon})$ as boundary degrees of freedom.  We argue that in this case, if we interpret $t\cdot V$ as part of the Hamiltonian, one cannot obtain consistent formulations giving the correct value of energy without fixing  (gauge) degrees of freedom.

We rewrite the boundary term in the canonical action as
\begin{equation}\label{SboundarySP3}
 S_{\text{SD,3}}^{\Delta} =-\int \md t\oint_{S_\Delta} \, (\mathcal{L}_t\psi_R + t\cdot V ) \,\leftidx{^2}{\epsilon} \, ,
\end{equation}
with the first term giving the required kinetic term, and we interpret the second term as a boundary part of the Hamiltonian
\begin{equation}\label{Ham3}
  H_{\text{SD,3}}= H_{\text{SD,1}} + \oint_{S_\Delta} \, (t\cdot V)\,\leftidx{^2}{\epsilon} \, .
\end{equation}
Its variation gives
\be
 \delta H_{\text{SD,3}}|_{S_\Delta}=  \oint_{S_\Delta} -\,\kappa_{(t)}\,\delta\eps+\delta(t\cdot V)\eps \,. \label{VarH3} 
\ee

On the other hand, 
\be\label{SymplStructSD3Var}
\Omega_{\text{SD,3}}(\de ,\de_t)|_{S_\Delta} =- \oint_{S_\Delta} \de_t\psi_R\,\de\,\eps - \de\psi_R\,\de_t\eps\, .
\ee
Differentiability of this Hamiltonian requires $\delta(t\cdot V)=A\,\delta\psi_R+B\,\delta\,\eps$, for some functions $A$ and $B$, and comparing (\ref{VarH3}) and (\ref{SymplStructSD3Var}) would give Hamilton's equations $\La_t\psi_R=\de_t\psi_R=\kappa_{(t)}-B$ and $\La_t\eps=\de_t\eps=-A$. Thus, for consistent equations we need a counter term such that
\[
\de H_{\text{CT,3}}=-\oint_{S_\Delta}\, \de(t\cdot V)\,\eps\,.
\] 
Taken together, these equations imply the counter term and $t\cdot V$  must now be functions of area: $H_{\text{CT,3}}=Fa_\Delta$ and  $t\cdot V=f(a_\Delta)$. So a consistent formulation necessarily requires fixing the gauge degrees of freedom of $t\cdot V$ and does not relate $\kappa_{(t)}$ to horizon area.
For the physical Hamiltonian flow $\de_t\psi_R=0$, we require a counter term such that
\[
\de H_{\text{CT,3}}= \oint_{S_\Delta} \,\kappa_{(t)}\,\delta\,\eps-\delta(t\cdot V)\,\eps \,,
\]
and we arrive at the same conclusion.

Finally, we observe again that in this extended space with boundary degrees of freedom $\psi_R$, we can get a consistent formulation without fixing the degrees of freedom of  $t\cdot V$, precisely by interpreting (the integral of) this term as giving a (non dynamical) boundary term for the symplectic structure. That is, by introducing $\psi_I$ as a non dynamical field with some initial conditions at $t_0$. This canonical phase space has symplectic structure (\ref{SymplStructSD1}) with only  the $\delta\psi_R$ variations treated as independent.
Again, calculations  identical to those for \textit{Canonical extension I}, give zero mass for the natural flow and the correct value for the physical dynamical flow $\de_t\psi_R=0$. This is the canonical phase space corresponding to \textit{Covariant extension III}.

\subsubsection{\bf{Canonical formulations with no boundary degrees of freedom}}

In this last case, there are no boundary degrees of freedom and the boundary term in the canonical action may be left as
\be
S_{\text{SD,4}}^{\Delta}=-\int \md t \oint_{S_\Delta} (\kappa_{(t)} + t\cdot V)\,\leftidx{^2}{\epsilon} \, ,
\label{CanonicalAction4}
\ee
and interpreted to contribute to the Hamiltonian.
There are no boundary contributions to the symplectic structure, and the differentiable functionals are defined imposing the Regge-Teitelboim type conditions \cite{Regge&Teitelboim}.  
The Hamiltonian is hence
\begin{equation}\label{Ham4}
  H_{\text{SD,4}}= H_{\text{SD,1}} + \oint_{S_\Delta} \, (\kappa_{(t)} + t\cdot V)\,\leftidx{^2}{\epsilon} \, ,
\end{equation}
and its variation has the boundary term:
\begin{equation}
\delta H_{\text{SD,4}}\vert_{S_\Delta} = 
 \oint_{S_\Delta}\, \big[\, (\delta\kappa_{(t)} + \delta  (t\cdot V)\bigr]\, \leftidx{^2}{\epsilon}
 \, .\label{VarHamBound4}
\end{equation}

Since $\Omega_{\text{SD,4}}$ does not have a boundary contribution, the boundary term (\ref{VarHamBound4}) must vanish or be canceled by a variaton of an appropriate counterterm, such that 
\be\label{VarHT3}
\delta (H_{\text{SD,4}}+H_{\text{CT,4}})\vert_{S_\Delta}=0 \, .
\ee

Here again, a similar analysis as in \textit{Canonical extension III} shows that we cannot get a consistent formulation without fixing the gauge degrees of freedom of $t\cdot V$.

A consistent formulation without boundary degrees of freedom and without fixing $t\cdot V$ is of course possible if one introduces $\psi_R$ and $\psi_I$, along with initial conditions for them at some time $t_0$. Once again, the boundary term (\ref{CanonicalAction4}) written as (\ref{CanSympl1}) implies the symplectic structure is (\ref{SymplStructSD1}), with boundary variations $\delta\psi$ determined by the bulk variations. 
Calculations are again the same as in \textit{Canonical extension I} with $\de_t\psi_R=0$.

In summary, our canonical analysis shows that we can get well defined canonical formulations by considering the boundary term $t\cdot\omega$ as contributing to either the Hamiltonian or the symplectic structure, and that we may consider the contribution to the symplectic structure as dynamical (due to boundary degrees of freedom) or not. On the other hand, our calculations show that the term $t\cdot V$ must always be interpreted as contributing to the symplectic structure to get a consistent canonical formulation, and furthermore that this contribution may be interpreted as dynamical (or not).

Let us emphasize that our analysis includes various aspects of the covariant and canonical phase space approaches, for spacetimes with a WIH as an internal boundary, that have not been noticed before. 
As we have mentioned earlier, in the covariant phase space formalism, only the case where $\psi_R$ and $\psi_I$ have been completely fixed has been considered. On the other hand, in the canonical approach originally only $\psi_I$ has been considered as a non-dynamical boundary degree of freedom. The purpose of this manuscript was to explore all possible extensions of the corresponding phase spaces and show how one can achieve the correspondence between different descriptions.


\section{Summary and Conclusions}
\label{sec:5}

We recapitulate what we have done. The main objective of this manuscript is to contribute to a better understanding of covariant and canonical Hamiltonian descriptions of spacetimes with an isolated horizon as a boundary, and the relation between them. Our treatment was based on a formulation of general relativity with a first order action, in self-dual variables. This choice was motivated by the relative simplicity of the resulting description. We were specially interested in the WIH contribution to the symplectic structure and the Hamiltonians of the theory. Our results are obtained for the special choice of gravitational variables in spacetimes with WIH as an internal boundary. There are many other contributions, based on different choices of bulk and boundary variables and with a variety of boundaries, as, for example \cite{Barbero&al,Freidel,B&F,B&F1}. Furthermore, the comparison of the boundary dynamics in covariant and canonical phase spaces has not been considered from the point of view that we wanted to explore in this manuscript. Let us also mention that recently the relation between the null infinity as a boundary of Penrose completition of an asymptotically flat spacetime and WIH has been discovered \cite{AshtekarSpeziale}, but it is not clear if one could straightforwardly apply the analysis that we performed here for WIH to the null infinity. Let us review our main results.  

Within the covariant approach, one can directly obtain the symplectic structure  which contains a boundary term at the horizon. To exhibit this explicitly however, requires the introduction of auxiliary, non-dynamical potential fields $\psi_R$ and $\psi_I$ at the horizon. These fields are defined so that their time derivatives match some of the degrees of freedom of the connection (equations (\ref{real_psi}) and (\ref{imaginary_psi})). To guarantee no new or extra degrees of freedom are introduced, initial conditions for the potentials are carefully fixed.
Using standard arguments one can further obtain an expression for the first law of horizon dynamics and for the horizon energy. 
There is some freedom in the exact choice of initial conditions for the potentials, but this ambiguity has no bearing on degrees of freedom or the Hamiltonian.
A key observation from this contribution is thus the understanding that the horizon boundary term in the symplectic structure does not signal the presence of boundary degrees of freedom at the horizon. The self-dual action does not lead to  covariant or canonical phase spaces with independent WIH boundary degrees of freedom. Nevertheless, the unavoidable contribution from variations of the auxiliary fields $\psi_R$ and $\psi_I$ to the symplectic structure, naturally suggests,  one may consistently consider these fields as independent degrees of freedom at the WIH.
We have explored this possibility here and shown that this is in fact viable.

There are three possible extensions of the covariant framework which result from lifting restrictions or initial conditions on each $\psi_R$, $\psi_I$, or both, and regarding the corresponding ambiguities introduced as true degrees of freedom.
Not surprisingly, in the two extensions considering $\psi_R$ as independent, the natural dynamical flow of this field, $\delta_t\psi_R=\mathcal{L}_t\psi_R$, has to be modified or frozen $\delta_t\psi_R=0$, if one wants to recover the correct value of energy and the first law.
In contrast, given that the ambiguities introduced by $\psi_I$ can be interpreted as special $U(1)$-gauge transformations of the transverse connection $V$, its natural dynamical flow $\delta_t\psi_I=\mathcal{L}_t\psi_I$ matches the flow determined by the bulk variables. An extension considering the gauge degrees of freedom encoded in $\psi_I$ as independent boundary degrees of freedom does recover the correct value of energy and the first law in a natural way.

In the canonical framework, the presence of a horizon boundary term in the canonical symplectic structure is revealed in a slightly different and subtle way. A straight forward reading or interpretation of the boundary terms arising from the 3+1 decomposition of the action, as contributing to the Hamiltonian, leads to an inconsistent or incomplete canonical formulation. The boundary term $\oint_{S_\Delta} (t\cdot V)\,\leftidx{^2}{\epsilon}$ has to contribute to the canonical symplectic structure and one has to introduce the non-dynamical field $\psi_I$ to do so.
There seems to be an ambiguity in interpreting the other boundary term $\oint_{S_\Delta} (\kappa_{(t)})\,\leftidx{^2}{\epsilon}$ as either contributing to the Hamiltonian or, by means of introducing $\psi_R$ as a non-dynamical field, to the symplectic structure. Although these choices lead to seemingly different symplectic structures,  both choices ultimately lead to the same Hamiltonian, giving the appropriate WIH mass and first law.

There is an exact correspondence between the original covariant phase space and the canonical phase space. The canonical symplectic structure (\ref{SymplStructSD1}) matches or more accurately, is $\bar{\Pi}$-equivalent to the covariant symplectic structure (\ref{pal_ss1}), with bulk-dependent variations $\delta_{1,2}\psi$ of the non-dynamical auxiliary fields $\psi=\psi_R+\psi_I$.
Furthermore, just like in the covariant case, one may construct three canonical extensions by regarding $\psi_R$, $\psi_I$, or both, as independent boundary degrees of freedom. These canonical extensions with extra boundary degrees of freedom, exactly match the three covariant extensions discussed. 
The  canonical symplectic structures are $\bar{\Pi}$-equivalent to the corresponding covariant symplectic structures. 
The natural Hamiltonian flow in each extended canonical phase space matches the natural dynamical flow on the respective covariant phase space, giving the same corresponding values for energy. 
Accordingly, here too, for the canonical extensions regarding $\psi_R$ as independent, one must modify or freeze the dynamical Hamiltonian flow $\delta_t\psi_R=0$ to get the `correct' or physical value of energy.
As already summarized above, there is an additional consistent canonical extension, resulting from the apparent ambiguity of interpreting the term involving $\kappa_{(t)}=t\cdot\omega$ as contributing to the Hamiltonian rather than to the symplectic structure. This extension does not seem to have a covariant analog.

It is most interesting that the particular extensions regarding $\psi_I$ as boundary degrees of freedom are very natural, in the sense that no  prescriptions or restrictions have to be added to derive the dynamical flow of the new degrees of freedom. These extensions amount to reinterpreting the (bulk) gauge degrees of freedom encoded in the term $t\cdot V$ as independent boundary degrees of freedom. Our construction and analysis gives further classical support to the quantization strategy followed in \cite{ABCK, ABK}.

The main lesson we can take from this analysis is that in both Hamiltonian methods the issues become subtle and one has to thread fine to arrive at correspondence between the two of them. We hope that this contribution can serve as motivation for further work on this important matter.



\begin{acknowledgments}
This work was in part supported by a CIC-UMSNH grant.
\end{acknowledgments}


\end{document}